\documentclass[aps,twocolumn,showpacs,pra]{revtex4}

\usepackage{exscale}
\usepackage{graphicx}
\usepackage{amsmath}
\usepackage{latexsym}
\usepackage{amsfonts}
\usepackage{amssymb}
\usepackage{times}

\usepackage{epsfig,pstricks}

\newcommand{\ketbra}[1]{\ensuremath{| #1 \rangle \langle #1 |}}
\newcommand{\braket}[2]{\langle #1 | #2 \rangle}
\def\trace{{\rm tr}\;}
\def\tr{{\rm tr}\;}
\def\CC{{\rm\kern.24em \vrule width.04em height1.46ex depth-.07ex
\kern-.30em C}}
\newcommand{\beq}{\begin{equation}}
\newcommand{\beqa}{\begin{eqnarray}}
\newcommand{\eeq}{\end{equation}}
\newcommand{\eeqa}{\end{eqnarray}}
\newcommand{\bra}[1]{\left\langle #1 \right |}
\newcommand{\ket}[1]{\left | #1 \right\rangle}
\def\trace{{\rm tr}\;}
\def\e{{\rm e}}
\newcommand{\eps}{\varepsilon}
\newcommand{\bigfrac}[2]{\mbox {${\displaystyle \frac{ #1 }{ #2 }}$}}

\renewcommand{\Re}{{\rm Re\;}}
\def\RR{{\rm
         \vrule width.04em height1.58ex depth-.0ex
         \kern-.04em R}}
\def\id{{\rm 1\kern-.22em l}}

\DeclareTextSymbolDefault{\i}{OT1} 

\begin{document}

\title{Estimating multipartite entanglement measures}

\author{Andreas Osterloh$^{1,2}$ and Philipp Hyllus$^{1,3}$}
\affiliation{$^1$Institut f{\"u}r theoretische Physik, Leibniz Universit{\"a}t Hannover,
Appelstr. 2, 30167 Hannover, Germany.\\
$^2$Fakult{\"a}t f{\"u}r Physik, Campus Duisburg, 
Universit{\"at} Duisburg-Essen, Lotharstr. 1, 47048 Duisburg, Germany.\\
$^3$BEC-INFM, Dipartimento di Fisica, Universit{\`a} di Trento, Via Sommarive 14, I-38050 Povo, Italy.}

\begin{abstract}
We investigate the lower bound obtained from experimental data of a 
quantum state $\rho$, as proposed independently by G\"uhne et al. and 
Eisert et al. and apply it to mixed states of three qubits. 
The measure we consider is the convex-roof extended three-tangle.
Our findings highlight an intimate relation to lower bounds 
obtained recently from so-called characteristic 
curves of a given entanglement measure. 
We apply the bounds to estimate the three-tangle present in recently performed
experiments aimed at producing a three-qubit GHZ state.
A non-vanishing lower bound is obtained if the GHZ-fidelity of the produced
states is larger than $3/4$. 
\end{abstract}


\maketitle

\section{Introduction}

Since entanglement has been recognized as a possible valuable resource for
quantum information processing \cite{nielsen}, its analysis, detection and 
quantification are major goals 
\cite{Plenio07,Horodecki07,AFO07,GuehnePR09}. 
Except for simple measures that are polynomial invariants of 
degree $2$, as e.g. the concurrence, already the calculation of that measure 
for a known mixed state poses a hard problem. In the laboratory, however, 
the states to work with are mixed states with major weight on a desired 
entangled state and additional uncontrolled
admixture of different states due to systematic or nonsystematic errors in 
the preparation process. Even though in priciple full state tomography can be
performed, it is experimentally increasingly expensive with the number of
qubits. 

It is for this reason that entanglement witnesses~\cite{Horodecki96} 
play an important role for experimental detection of entanglement. 
Witness operators are constructed such to have
negative expectation values only on states that carry a specific class
of entanglement; all states that do not belong to this class have positive
expectation value of the witness. Then, a negative expectation value 
implies that the mixed state $\rho$ carries the specific entanglement class 
detected by the witness. Recently, also quantitative estimates have been 
obtained for a variety of entanglement measures from (e.g. experimental) 
values for one or more 
witness operators~\cite{BrandaoPRA05,Guehne07,Eisert07,Guehne08},
and for non-linear combinations of expectation 
values~\cite{AudenaertNJP06,Wunderlich09a,Wunderlich09b},
by using methods of convex optimization~\cite{BoydBook04}.
Further lower bounds have been obtained using a different
approaches in a similar context~\cite{VerstraetePRL02,ChenPRL05,Breuer06,Vicente07,Datta07,Zhang07}. 
A connection of the problem of entanglement estimation from 
uncomplete information to Jaynes principle has been discussed already 
in~\cite{HorodeckiPRA99}.

Here, we investigate the tight lower bound for convex-roof extended measures
as introduced in Refs.~\cite{Guehne07,Eisert07} for measures quantifying true 
multipartite entanglement in order to compare them
with lower bounds obtained from different premises~\cite{KENNLINIE}.
The computation of the lower bound involves a supremum over some
parameters and an infimum over pure states.
One of our central observations is that the bound can be alternatively
computed in a way which involves an infimum over pure states only,
followed by the convexification of the resulting function.
This observation establishes a connection
between the methods from Refs.~\cite{Guehne07,Eisert07} and 
Ref.~\cite{KENNLINIE}.

To this end we focus on the three-tangle, where a specific analytic solution of
the convex roof can be used as a benchmark.

The work is laid out as follows. In the next section we briefly sketch the 
estimation method from witness operators as proposed in Ref.~\cite{Guehne07,Eisert07}
and introduce the three-tangle and general multipartite entanglement witnesses.
We then turn to the application of the method
in Section~\ref{sec:application} using expectation values of 
entanglement witnesses designed to detect true three-partite 
entanglement. We first consider a specific class of 
rank two mixed three qubit states 
for two different witness operators. The estimation using the data of the first 
witness is directly related to the results of Ref.~\cite{KENNLINIE}.
Then, we compute a lower bound for an important witness, which 
can directly be used to estimate the three-tangle produced in recent experiments.
Finally, we report on further interesting observations as the effect 
of common symmetries of the Witness operators and the entanglement measure in 
Section~\ref{sec:observations}, where we also sketch a proof for
the possible reduction to the related problem for pure states.
Section~\ref{sec:conclusions} contains our conclusions.

\section{Basic concepts}
\label{sec:basics}

\subsection{Lower bound on entanglement}\label{sec:lb}

We consider the following situation:
assume that a source can be described by the 
(unknown) density matrix $\rho$ and that 
$K$ expectation values $w_k=\trace[\rho {\cal W}_k]$ are measured
and then collected in a vector $w$.
The operators ${\cal W}_k$ are further assumed to be witness operators.
The lowest value of the entanglement of the state $\rho$ consistent
with the measurement results is formally given by the solution of the problem
\beq
	\inf_\rho E(\rho)\big|_{\trace[{\cal W}_k\rho]=w_k}\ .
	\label{eq:mixedproblem}
\eeq
In Refs~\cite{Guehne07,Eisert07} it has been shown that the solution of
this optimization problem can be approximated from below with affine functions by 
$\epsilon(w)=\sup_r\Big(r\cdot w-{\hat E}(\sum_k r_k {\cal W}_k)\Big)$,
where $\hat E({\cal W})=\sup_\rho\big(\trace[\rho{\cal W}]-E(\rho)\big)$ is the Legendre 
transform of $E$.
It was further shown that for pure state entanglement measures extended {\em via}
the {\em convex roof construction} as 
\beq
	E(\rho):=\inf_{\rho=\sum_i p_i \ketbra{\psi_i}}\sum_i p_i E(\ket{\psi_i}),
	\label{eq:convexroof}
\eeq
where $\sum_i p_i=1$ and $p_i\ge 0$, the Legendre transform can be computed
by optimizing over pure states only, leading to
\beq\label{supinf}
	\epsilon(w)=\sup_r\inf_{\ket{\psi}}
	\Big(\sum_{k=1}^K r_k (w_k-
	\bra{\psi} {\cal W}_k \ket{\psi})+E(\ket{\psi})\Big).
\eeq
This bound is tight due to the convexity of the problem \cite{Guehne07,Eisert07}.

A key observation is that Eq.~\eqref{supinf} is the dual problem to
the minimization of $E(\psi)$ on pure states subject to the witness conditions. 
It is therefore solved by an approximation of the pure state problem
\beq
	\inf_\psi E(\psi)\big|_{\bra{\psi}{\cal W}_k\ket{\psi}=w_k}
	\label{eq:pureproblem}
\eeq
from below with affine functions
Hence the result coincides with the function convex 
hull of the pure state problem~\footnote{In order to make this point more clear, we present a derivation
of the bound in the language of the Lagrange dual 
problem~\cite{BoydBook04}. The Lagrange dual function to 
problem~(\ref{eq:mixedproblem}) is defined as 
$g(\vec r)=\inf_\rho E(\rho)+\sum_k r_k (\nu_k-\trace[{\cal W}_k\rho])$,
where the Lagrange multipliers $r_k$ are real numbers. This is a lower bound to the optimal solution 
of the first problem since if $\tilde\rho$ is a state fulfilling all the constraints 
then the terms proportional to $\nu_k$ vanish. 
For convex-roof extended measures, we obtain
$\inf_\rho \big\{[E(\rho)+\sum_k r_k (w_k-\trace[{\cal W}_k\rho])\big\} = \inf_{\{p_i,\ket{\psi_i}\}} \sum_i p_i \Big(E(\ket{\psi_i})+\sum_k r_k (w_k- \bra{\psi_i}{\cal W}_k\ket{\psi_i}\Big)\ge \inf_{\ket{\psi}} \big\{E(\ket{\psi})+\sum_k r_k (w_k-\bra{\psi}{\cal W}_k\ket{\psi})\big\}$,
where $\rho=\sum_i p_i \pi_{\psi_i}$,  $\sum_i p_i=1$, $p_i\ge 0$,
and $\pi_\psi:=\ket{\psi}\bra{\psi}$ is the 
projector on the state $\ket{\psi}$. The fact that $E(\rho)$ 
is defined via the convex-roof extensions
from its value on pure states enters in the first inequality.
The optimal lower bound is then obtained by taking the supremum over all Lagrange multipliers
$\vec{r}$, leading to Eq.~(\ref{supinf}). The claim follows from the fact
that the final expression above
is the Lagrange dual function of the pure state 
problem~(\ref{eq:pureproblem}).
}.
In other words, in order to solve the problem~(\ref{eq:mixedproblem})
for convex-roof extended entanglement measures,
we could alternatively solve the problem~(\ref{eq:pureproblem})
and convexify the resulting function. In cases where $K$ is 
small, this could reduce the computational cost.
This point is illustrated in Section~\ref{sec:restricted},
where we explicitly solve~(\ref{supinf}) and~\eqref{eq:pureproblem} 
for a restricted situation, and the equivalence of both problems
is highlighted explicitly. The origin of this equivalence is further investigated in 
Section~\ref{pure-cond} for arbitrary $K$.

The entanglement measure we consider is the
three-tangle and its square; it distinguishes the two classes of global entanglement
for three qubits: W and GHZ.

\subsection{Three-tangle}

The three-tangle of a general 
pure three qubit state expanded in a product basis $\ket{\psi}=\sum_{i,j,k=0}^1\psi_{ijk}\ket{ijk}$
is given by~\cite{Coffman00}
\beqa\label{def_3tangle}
\tau_3 &=& 4\ |d_1 - 2d_2 + 4d_3|\\
  d_1&=& \psi^2_{000}\psi^2_{111} + \psi^2_{001}\psi^2_{110} + \psi^2_{010}\psi^2_{101}+ \psi^2_{100}\psi^2_{011} \nonumber \\
  d_2&=& \psi_{000}\psi_{111}\psi_{011}\psi_{100} + \psi_{000}\psi_{111}\psi_{101}\psi_{010}\nonumber \\ 
    &&+ \psi_{000}\psi_{111}\psi_{110}\psi_{001} + \psi_{011}\psi_{100}\psi_{101}\psi_{010}\nonumber \\
    &&+ \psi_{011}\psi_{100}\psi_{110}\psi_{001} + \psi_{101}\psi_{010}\psi_{110}\psi_{001}\nonumber \\
  d_3&=& \psi_{000}\psi_{110}\psi_{101}\psi_{011} + \psi_{111}\psi_{001}\psi_{010}\psi_{100}\nonumber .
\eeqa
It can be extended to mixed states via the convex-roof construction~(\ref{eq:convexroof}).

The three-tangle is non-vanishing on true 
3-partite entangled states only and vanishes on any 
bi-separable state such as $\ket{\psi_{AB}}\otimes\ket{\phi_C}$
for the parties $A$, $B$, and $C$, but also for W states
introduced below.
Also the convex-roof extended measure has this property
for mixtures of bi-separable states, 
possibily for different partitions
of the parties.
Summarizing, it distinguishes the two classes of 
three-partite entanglement that are inequivalent under stochastic 
local operations and classical communications (SLOCC) \cite{Duer00}. 
The representative of one class is the W-state 
\beq
	\ket{\rm W}=\frac{1}{\sqrt{3}}(\ket{001}+\ket{010}+\ket{100})
\eeq
and the representative of the other class is the GHZ state \cite{Greenberger89}
\beq
	\ket{\rm GHZ}=\frac{1}{\sqrt{2}}(\ket{000}+\ket{111}).
\eeq
The three-tangle vanishes on all states of the W-class
while it is non-vanishing for states of the GHZ-class.
This classification can be extrapolated to mixed states
in the sense that a mixed state belongs to the W-class if it has
a decomposition into pure states exclusively out of the 
W-class~\cite{Acinbe}. In analogy, a mixed state
belongs to the GHZ-class if it has a decomposition into 
pure states of the GHZ-class.

\subsection{Multipartite entanglement witnesses}

Multipartite entanglement witnesses can be constructed as~\cite{Bourennane04}
\beq
	{\cal W}=\alpha\id-\ketbra{\phi},
	\label{Eq:wit}
\eeq
where $\alpha=\max_{\ket{\text{Zero}}}|\braket{\phi}{\text{Zero}}|^2$
is the maximal overlap of $\ket{\phi}$ with any state for which a given
entanglement measure $E(\ket{\text{Zero}})=0$. In the case that global 
entanglement is concerned, $\ket{\text{Zero}}$ would be all biseparable states,
whereas also $W$-states were included if only states with a GHZ component 
are of interest \cite{Coffman00,Duer00,Acinbe}.
We will highlight the latter situation and use the three-tangle, $E:=\tau_3$,
since it is non-zero only for states of the GHZ class. 

For such witnesses, Eq.~(\ref{supinf}) becomes
\beq
	\sup_r \left\{-\sum_{k=1}^K r_k \bra{\phi_k}\rho\ket{\phi_k} + 
	\inf_{\ket{\psi}} \big(	E(\ket{\psi}) + \sum_{k=1}^K r_k |\braket{\phi_k}{\psi}|^2 \big)\right\}.
	\label{eq:projwitsupinf}
\eeq
This tells us that the trivial part $\alpha\id$ of the witness is irrelevant
for the estimation.
Hence, it does not seem to be essential that the expectation
values are measured with respect to witnesses; an operator
which projects onto a suitable subspaces seems to
be sufficient \cite{Guehne08}. 

For a single witness of this form 
it is straight forward to check that $r\geq 0$ leads to $\epsilon(w)=0$;
indeed, in this case the infimum over all states $\ket{\psi}$ would be
 those states perpendicular to $\ket{\phi}$ for which $E(\psi)=0$.
This is always possible for a single witness.
Another peculiar case is when $\bra{\phi}\rho\ket{\phi}=0$. 
In this case the estimate is independent of the density matrix under
consideration and hence can only be zero. 
This corresponds to a bad choice for the witness operator.
For the three-tangle, choosing $\ket{\phi}$ outside the GHZ class
also leads to $\epsilon(w)=0$ since then the infimum in 
(\ref{eq:projwitsupinf}) is reached by choosing 
$\ket{\psi}=\ket{\phi}$ for $r<0$.   

The above reasoning has interesting implications. 
First, $r$ can be taken to be negative
for a single witness operator detecting the class of entanglement 
measured by $E$.
Second, it will admit to relax the restriction for ${\cal W}$ to being a witness, 
and a wider class of observables might lead to a reasonable 
estimation of $E$. 
In these cases, $r$ can possibly also be positive~\footnote{
A trivial example is the ``anti-witness'' $\bar{\cal W}=-{\cal W}$ of 
a witness ${\cal W}$.}. 
For more than one witness, the situation is more
complicated and then even observables that are not useful at all for
detecting entanglement could possibly be useful for improving
the lower bound.

\section{Application}
\label{sec:application}

We illustrate the results of the previous Section by explicitly solving 
the optimization problem~\eqref{supinf} for special cases where knowledge
about the state is assumed to be given and by showing explicitly
that the result coincides with the one obtained by solving 
problem~(\ref{eq:pureproblem}) and convexifying the solution.
Then, we numerically 
solve the general problem for a general witness which allows
to estimate the three-tangle from experiments where the GHZ state
was produced and the fidelity $\bra{\rm GHZ}\rho_{\rm exp}\ket{\rm GHZ}$
was estimated for the produced state $\rho_{\rm exp}$.

\subsection{Restricted optimization}
\label{sec:restricted}

We will demonstrate the Legendre transform at work on a simple example,
where the Hilbert space for minimization with respect to $\psi$ is restricted
to the range of the actual density matrix. This amounts to an idealized
situation where the experimenter knows precisely which states are possibly
produced in the laboratory to be part of the resulting mixed state.   
The specific example is about the three-tangle of the three qubit mixed states
\beq
	\rho(p)=p\ \pi_\text{GHZ}+(1-p)\ \pi_\text{W}\; ,
	\label{eq:teststate}
\eeq
for which the analytic convex roof is known~\cite{LOSU}. 
Here, $\pi_{\text{GHZ}}$, $\pi_{\text{W}}$ are the projectors onto 
the $\ket{\text{GHZ}}$ and $\ket{\text{W}}$ state, respectively.
The vectors in the range of $\rho(p)$ are given by
\beq
	\ket{Z(q,\phi)}=\sqrt{q}\ket{\text{GHZ}}-e^{i\phi}\sqrt{1-q}\ket{\text{W}}
	\label{eq:restrictedstates}
\eeq
with its three-tangle given by
\beq
	\tau_3(q,\phi):=\tau_3(Z(q,\phi))=\Big|q^2-\frac{8\sqrt{6}}{9}\sqrt{q(1-q)^3}e^{3i\phi}\Big|.
	\label{eq:tau3restricted}
\eeq
In the following, we want to apply the method of 
Refs~\cite{Guehne07,Eisert07} 
in order to get lower bounds for the three-tangle
of general states $\varrho$ in the GHZ-W subspace,
using two qualitatively different single witnesses
by performing the optimization of Eq.~(\ref{supinf}).
We the compare the result with that obtained by
solving the problem~(\ref{eq:pureproblem}) and by
convexifying the solution.

\subsubsection{GHZ witness}\label{single-wit}

\begin{figure}
  \includegraphics[width=.44\textwidth]{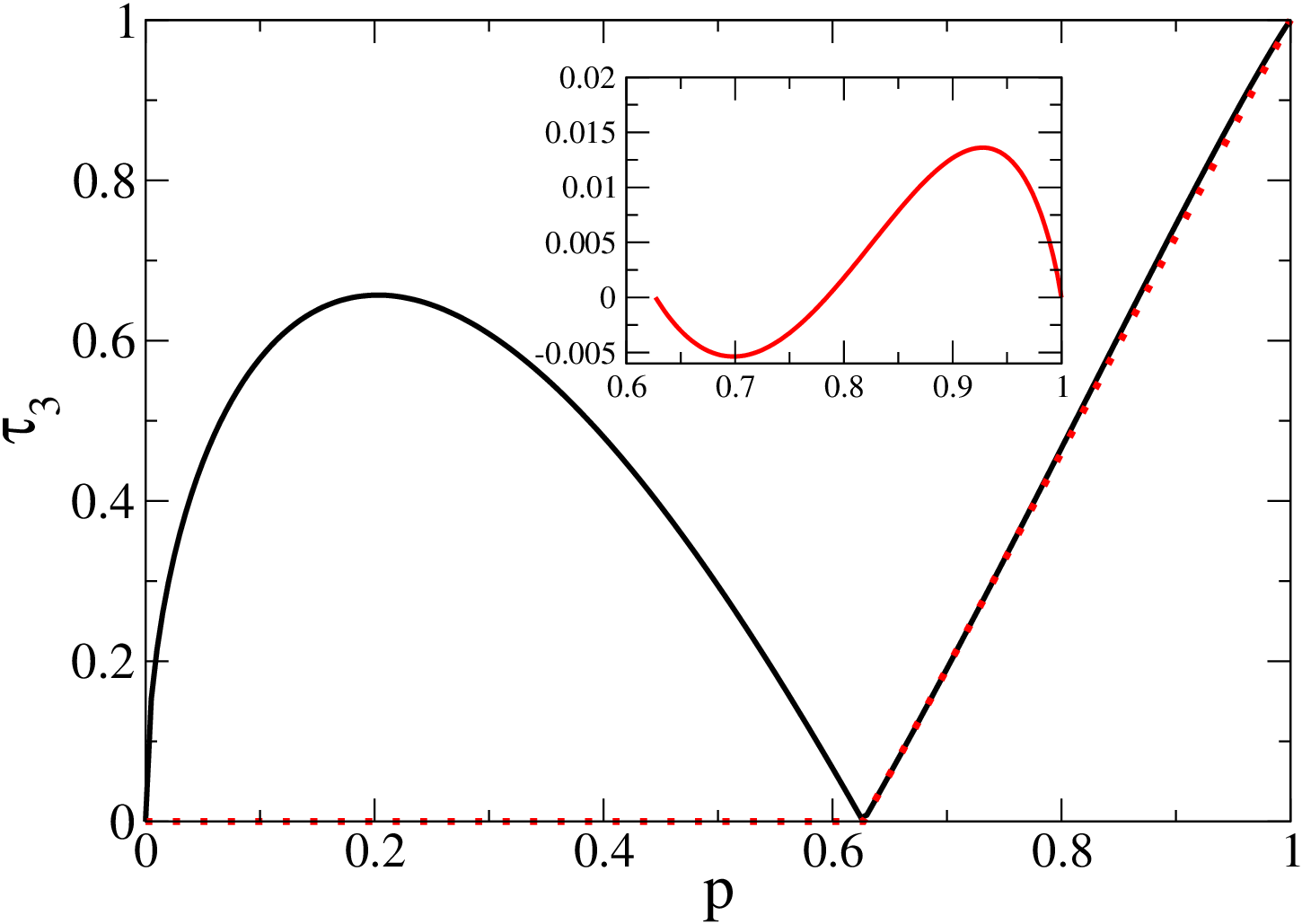}
 \caption{(Color online)
The graph shows how $\tau_3(q,0)$ varies with $q$. The
 graph reaches $0$ at $q=0$ and 
$q_0\approx 0.627$. The red dotted curve is the convex-roof $\hat{\tau}_3(p)$
for the mixed state~\eqref{eq:teststate}.
{\em Inset:} Deviation of $\tau_3(q,0)$ from the straight line going 
from $(q_0,\tau_3(q_0))$ to $(1,1)$ in the interval $[q_0,1]$. 
Note that the behaviour of $\tau_3(q,0)$ changes from convex to concave.}
  \label{fig:tau3}
\end{figure}

The first witness we would like to consider is
\beq
	{\cal W}_\text{GHZ}=\alpha\id-\pi_{\text{GHZ}}.
	\label{eq:GHZwit}
\eeq
For $\alpha=\frac{1}{2}$,
this is a witness for multipartite entanglement, while for $\alpha=\frac{3}{4}$,
it detects states of the GHZ class only \cite{Acinbe}. For this witness,
$w=\tr[{\cal W}\varrho]=\frac{1}{2}-p$, where we defined 
$p\equiv \bra{\phi}\varrho\ket{\phi}$. 
We perform the infimum with respect to $\ket{\psi}$ only over
the restricted set of states $\ket{Z(q,\phi)}$ in the range of 
$\varrho$. This is useful for obtaining a bound
on the tangle of the rank 2 state $\varrho$ and relies
on the assumption that we know the subspace that $\varrho$ lives
in.  As we will show below, 
knowing the expectation value of the witness~(\ref{eq:GHZwit})
corresponds to knowing $q$ in Eq.~(\ref{eq:restrictedstates}).
Hence the only free parameter in the pure state 
problem~(\ref{eq:pureproblem}) is the phase $\phi$,
which is minimized for $\phi=0$ \cite{LOSU}. By the argument 
of Section~\ref{sec:basics}, the lower bound to the 
problem is then given by the function convex hull $\hat{\tau_3}(q)$ of 
$\tau_3(q,0)$ (see Eq.~\eqref{eq:tau3restricted} and Fig.~\ref{fig:tau3}), 
corresponding to the convexified solution of problem~(\ref{eq:pureproblem}).
It has emerged as so-called {\em characteristic curve} 
from a different approach to obtain lower bounds on entanglement measures 
pursued in Ref.~\cite{KENNLINIE}.

\begin{figure}
  \includegraphics[width=0.44\textwidth]{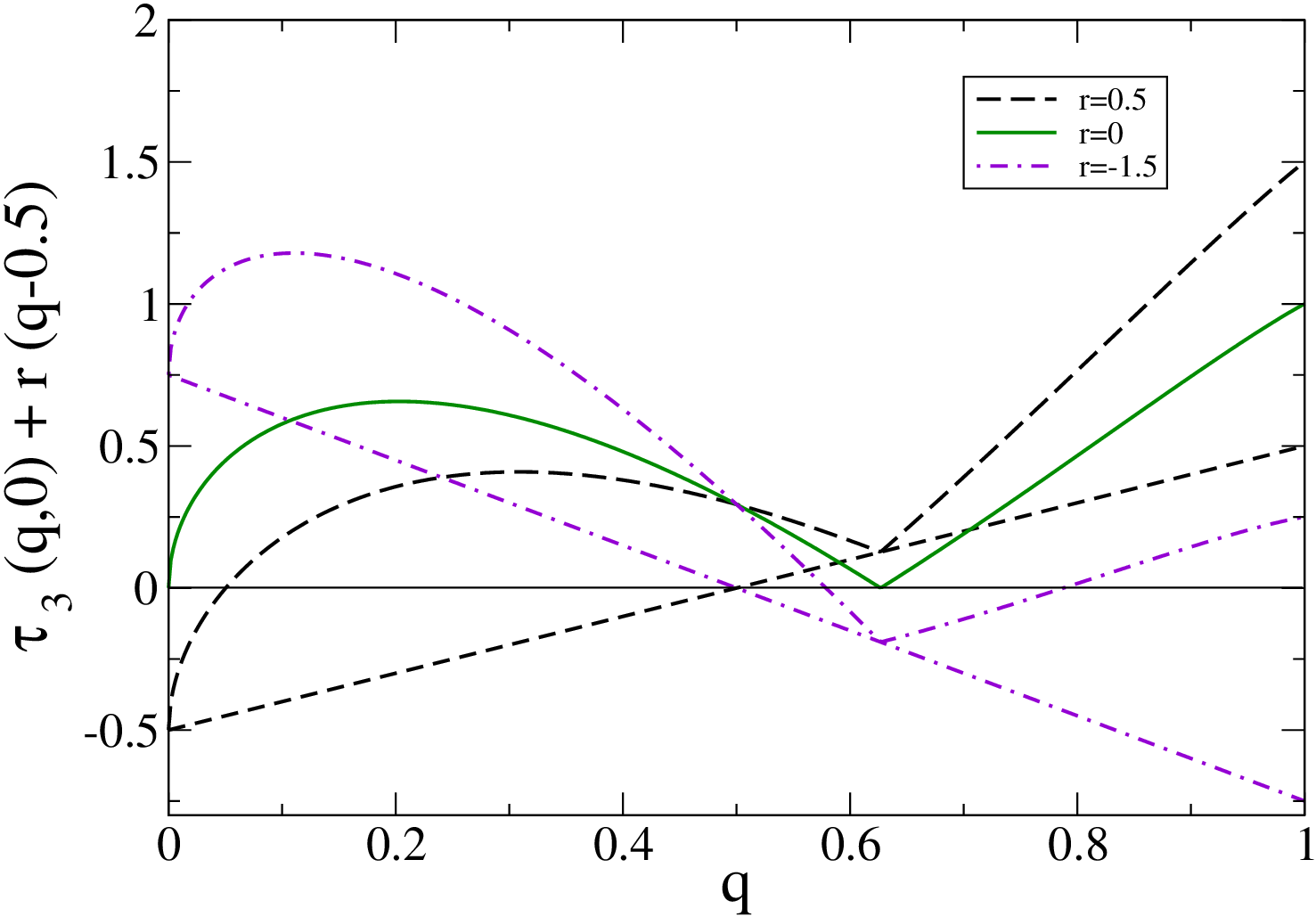}\\
\vspace*{11mm}
  \includegraphics[width=0.44\textwidth]{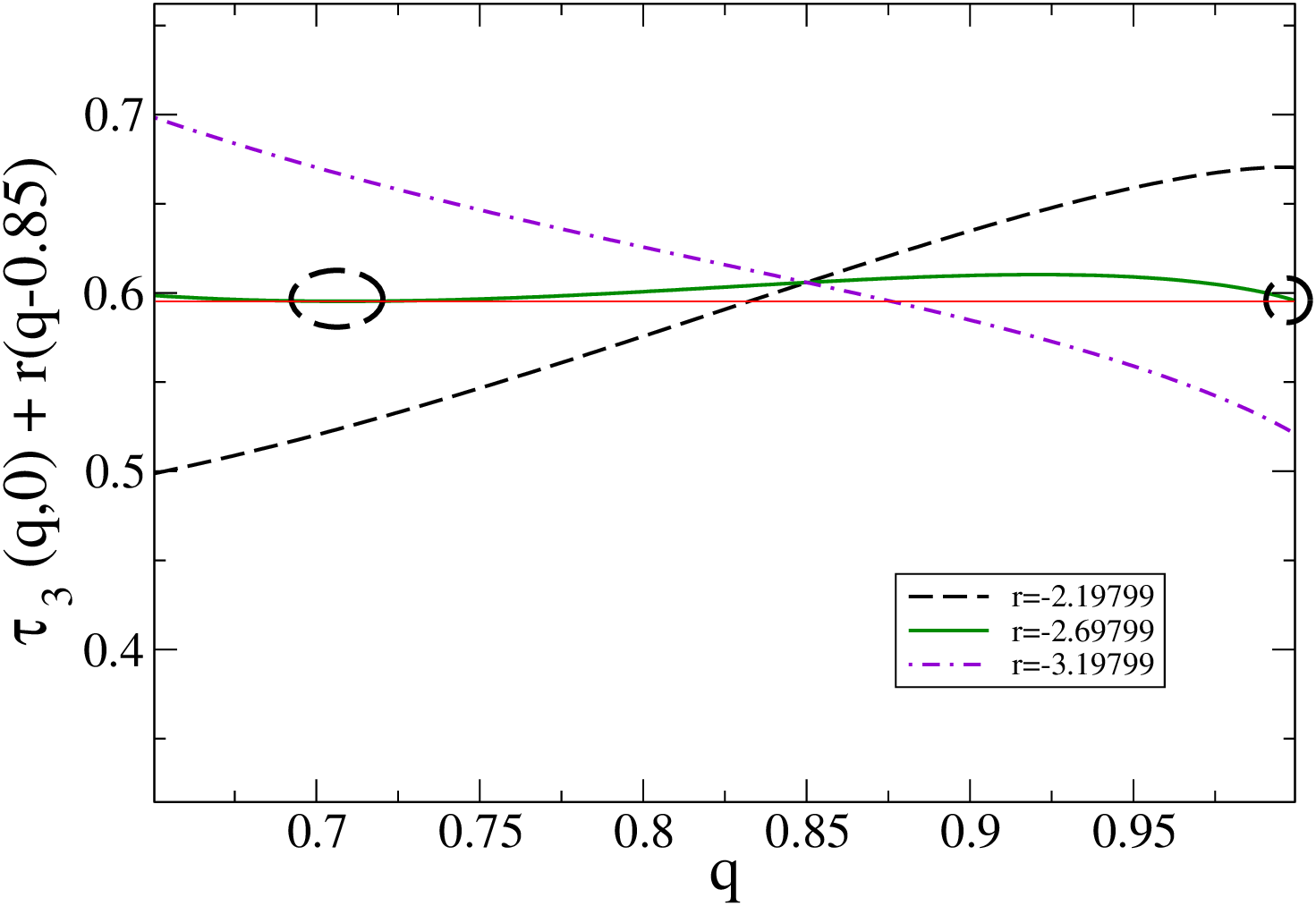}
\caption{(Color online) Here, we show the seesaw argument for the three-tangle.
The {\em upper panel} shows the situation for $p=0.5$ (corresponding to region (i)).
Besides the optimal value $r=0$ (green full line), 
one positive and one negative $r$ value
has been considerd; the largest absolute minimum of the tilted function
is clearly zero. The {\em lower panel} considers 
$p=0.85$ (corresponding to region (iii)). Here the optimal value $r=r_1$ is 
shown (green full curve) next to two close but deviating values.
\label{fig:seesaw}
}
\end{figure}

Let us compare this result to the problem
problem~(\ref{supinf}) which can be solved explicitly in this simple case. 
We have to determine
\beq\label{condition}
	\epsilon(w)=\sup_{r}\inf_{q,\phi}\Big(r(q-p)+\tau_3(q,\phi)\Big)
\eeq
and we show in what follows that 
$	\epsilon(w)=
	\hat\tau_3(p)\; $
is obtained also in this case.
As mentioned above, the infimum over $\phi$ is always 
obtained for $\phi=0$ so that we have to optimize 
over $q$ only.
$\tau_3(q,0)$ is plotted in Fig.~\ref{fig:tau3}.
Graphically, optimization in Eq.~\eqref{condition} means to 
tilt the $x$-axis about the fixpoint $(x,y)=(p,0)$, resulting
in a curve with local minima for each tilting slope $r$. 
The largest of these minima 
is the estimated lower bound for the entanglement measure, here $\tau_3$.
This {\em seesaw} argument is shown for two cases in Fig.~\ref{fig:seesaw}.

As already discussed earlier, assuming $r>0$ leads to an infimum at $q=0$ 
for all $p\in [0,1]$ and we are left with the trivial bound 
$\epsilon(w)=\sup_r(-rp)=0$.
So we take $r<0$ and focus at the infimum 
$\inf_{q}\big(-|r|q+\tau_3(q,0)\big)$,
formally looking for local extrema in 
intervals where $\tau_3(q)$ is twice differentiable. 
The differential minimum condition is then
\beqa\label{eqs:cond0}
|r|&=&\dot{\tau}_3(q,0) \\
\ddot{\tau}_3(q,0)&>& 0\; .
\label{eqs:cond}
\eeqa
The second condition simply demands that ${\tau}_3(q,0)$ be convex
(see the discussion in Section~\ref{pure-cond} for the general case).
Where ${\tau}_3(q,0)$ is concave, the infimum is found at the border
of some interval, which has to be determined.
We have to find the global infimum and three cases have to be
treated separately.
\begin{figure}
  \includegraphics[width=.44\textwidth]{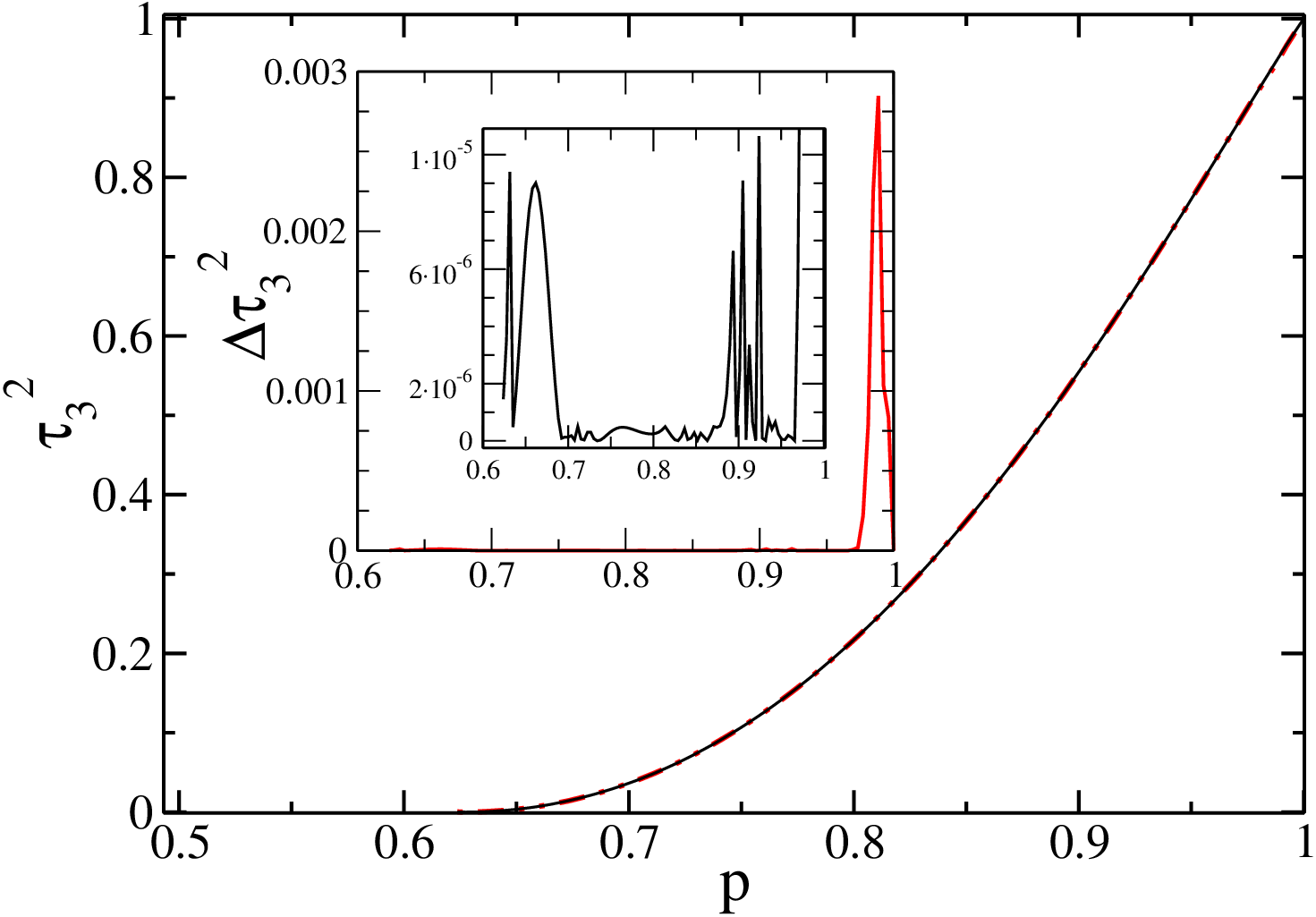}
 \caption{(Color online) The estimate for $\tau_3^2$ from a diagonal witness (red curve)
together with the square of the corresponding characteristic curve (black). 
It can be seen only in the inset that very close to $p=1$ this 
characteristic curve still fails to be convex; 
the lower bound correctly gives the function convex hull.
The small inset zooms into the region, where the 
characteristic curve fails to be convex. It shows the
numerical error, which is two orders of magnitude below this
deviation.}
  \label{fig:tau3estimate-diagonal}
\end{figure}

Unregarded the value of $p$ there are three separate regions for the 
variable $r$: 
(i) $0>r>r_0$, (ii) $r_0>r>r_1$, and
(iii) $r_1>r$ with $r_0$, $r_1$ to be specified below. 
The three cases can be understood graphically from Fig.~\ref{fig:tau3}.
In region (i), the infimum is taken at 
$q_0=\frac{4\sqrt[3]{2}}{3+4\sqrt[3]{2}} \ \approx\ 0.627$ 
until the slope of the function $-|r|q+\tau_3(q,0)$ (for $q\ge q_0$) 
at $q=q_0$ vanishes. This happens for 
$r_0=-\dot{\tau}_3(q_0,0)\approx -2.52$. 
In region (ii), the 
function $-|r|q+\tau_3(q,0)$ has a unique local minimum at some 
position $q_r\in[q_0,q_1]$
given by Eq.~\eqref{eqs:cond0}, where $q_1$ is determined below. 
This local minimum is also the global minimum.
We see that $-|r|q+\tau_3(q,0)$ is convex in this region, when zooming
into the plot, as shown in the right panel of Fig.~\ref{fig:tau3}.
Finally, $r_1$ is the value for $r$ below which the infimum is located 
at $q=1$. 
The values of $r_1$ and $q_1$ can be computed as follows: at $r=r_1$ 
the infimum is reached both at $q_1$ and at $q=1$. 
We therefore have $r_1 q_1+\tau_3(q_1,0)=r_1+\tau_3(1,0)$, 
and hence \mbox{$r_1=-\frac{1-\tau_3(q_1)}{1-q_1}$}.
Also at this point, the slope of $\tau_3(q,0)$ equals that of the straight
line connecting the points $(q_1,\tau_3(q_1,0))$ and $(1,1)$, hence
$r_1=-\dot{\tau}_3(q_1)$.
Both conditions determine 
$q_1=\ \frac{1}{2}\ +\ \frac{3}{310}\sqrt{465}\ \approx\ 0.70868$.
Reinserting $p$, we are left with the following optimization problems 
in the three regions:
(i) $\sup_r r(q_0-p)$, (ii) $\sup_r r(q_r-p)+\tau_3(q_r,0)$ ($q_r\in[q_0,q_1]$),
while in region (iii), the infimum is given
by $-|r|+1$, and $\epsilon(w)=\sup_r(-|r|(1-p)+1)=1-|r_1|(1-p)$. This curve is
the straight line connecting the points $(q_1,\tau_3(q_1))$ and $(1,1)$,
as already mentioned before.

The question we address now is, which of the above depicted three regions 
corresponds to a given $p$.
For $p\le q_0$, choosing $r$ from region (i) yields $\eps(p)=0$ since
$r<0$ and $q_0-p>0$. In regions (ii) and (iii) the infimum is at 
negative values (see Fig.~\ref{fig:seesaw}).
For $q_0<p\le q_1$, we find $\epsilon_{(i)}=r_0(p-q_0)$ in region (i), 
and $q_r=p$ in region (ii), 
hence $\epsilon_{(ii)}=\tau_3(p,0)$. 

In region (iii) the infimum
is at negative values again. Due to the convexity of $\tau_3(p,0)$ in
$p\in[q_0,q_1]$,  $\tau_3(p,0)>r_0(p-q_0)$, 
and hence $\epsilon(w)=\tau_3(p,0)$.
Finally, for $q_1<p\leq 1$ the supremum of the infima is located 
in region (iii), and $\epsilon(w)= 1-|r_1|(1-p)$.
Altogether, this yields $\epsilon(w)=\hat{\tau}_3(q)$ 
as claimed above. 

This is further illustrated by numerically applying both methods to 
$\tau_3^2$~\footnote{Whereas the discussion in the text
is done for the three-tangle itself, the numerical analysis has been performed for
the square of the three-tangle. The motivation is two-fold: first, it avoids square roots
and hence leads to a continuous derivative, which simplifies the numerical optimization
routine; second, it reduces the interval in which the characteristic curve is concave
(this can be seen by considering the second derivative of a 
positive semi-definite function $f(x)$ exponentiated with $n>1$:
d$^2 f^n/$d$x^2=nf^{n-1}f''+n(n-1)f^{n-2}(f')^2$, where $f'=$d$f/$d$x$. Since $f\ge 0$ holds,
d$^2 f^n/$d$x^2\ge 0$ wherever $f''\ge 0$).
It must be kept in mind though that $\widehat{\tau_3^2}\geq (\hat{\tau_3})^2$, and that
$\sqrt{\widehat{\tau_3^2}}$ is not convex.},
as shown in Fig.~\ref{fig:tau3estimate-diagonal}. 
The region, where the function convex hull must be applied is 
clearly distinguished from the numerical uncertainty.

Quite generally, lower bounds can be
obtained from the characteristic curve of a given entanglement 
measure~\cite{KENNLINIE}.
Here this lower bound coincides with that bound obtained from the approach in
Refs~\cite{Guehne07,Eisert07}, and even gives the analytic convex roof.
In this sense, the bound is tight.

\subsubsection{Off-diagonal witness}

Clearly, the previous discussion marks an ideal situation in 
that the infimum over the admissible states $\psi$ was easy to handle.
As already mentioned in Section~\ref{sec:basics}, it is essential for the 
operator ${\cal W}$ to
have a nontrivial overlap with the set of density matrices of interest.
With {\em nontrivial} we mean that $\trace\,\rho\, {\cal W}\neq const$. 
For GHZ-W mixtures it implies that $W$ must be able to distinguish 
both states. 
Consequently, 
${\cal W}'=-\alpha \ket{W}\bra{W}-\beta\ket{\rm GHZ}\bra{\rm GHZ}$ 
can be employed 
(unless $\alpha=\beta$), but also ${\cal W}''=-\ket{111}\bra{111}$,
both  without 
changing the resulting lower bound as obtained above for the particular 
GHZ witness ${\cal W}_{\text{GHZ}}=\frac{1}{2}\id-\pi_{\text{GHZ}}$. 
We emphasize however that this is due to the restricted Hilbert space 
we consider here. To point it differently: more ab initio information
about the state admits more freedom to the observables in order to get 
sensible lower bounds.

A more generic situation occurs when we admit for {\em off-diagonal} 
operators like
\beq\label{schief}
{\cal W}_{\rm skew}=-\pi_{\rm GHZ}
-\omega \ket{\rm GHZ}\bra{W}-\omega^*\ket{W}\bra{\rm GHZ}\; .
\eeq

\begin{figure}
\begin{center}
  \includegraphics[width=0.44\textwidth]{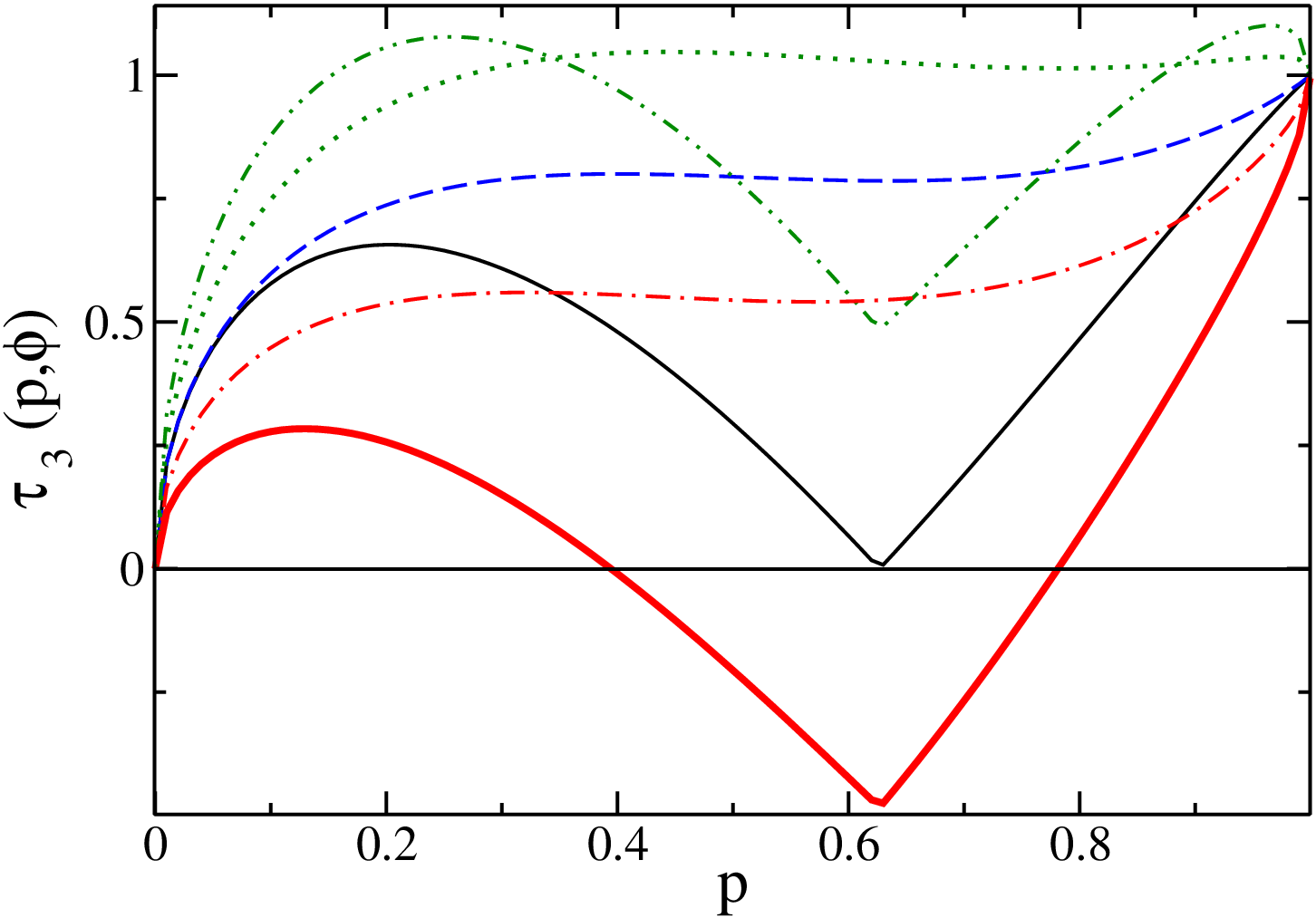}
\end{center}
 \caption{(Color online) For visualizing how the estimated lower bound is decreased
by the off-diagonal part of the witness~\eqref{schief}, we assume the term 
$\sqrt{q(1-q)}\cos\phi$ to appear without $r$ as a prefactor for a moment. 
It then simply modifies the characteristic curve as shown here for
$\omega=1$. The function
convex hull of the lowest envelopping curve (thick full red line) 
would then constitute the lower bound estimate from the off-diagonal witness.
However, this estimate would still lead to the correct value $1$ at $p=1$.
Even this is spoiled when the contribution from the off-diagonal part 
of the witness is actually $r$-dependent, as is the case here.}
\label{roughargument}
\end{figure}

We obtain
\beq\label{condition:skew}
	\epsilon(p)=\sup_{r}\inf_{q,\phi}\Big(r(q-p+2\sqrt{q(1-q)}\Re{\omega \e^{i\phi}})+\tau_3(q,\phi)\Big).
\eeq
In this case, the infimum cannot be taken separatly for $q$ and $\phi$.
It is clear from the specific situation that the off-diagonal part would
result in a decrease of the lower estimate of the three-tangle 
(see Fig.~\ref{roughargument}). 
The fact that $r$ is a prefactor of $\sqrt{q(1-q)}$ 
in the function to be optimized leads
to the feature that even the estimate at $p=1$ is below the exact value $1$.
We analyze this case in detail because it exhibits relevant features 
an improper choice for the witness might have. 
This type of experimental 
``error'' might be due to e.g. incomplete knowledge about relative phases 
in the state under consideration or a not perfectly symmetric setup for the 
production of e.g. a GHZ state. The restriction to a two-dimensional 
Hilbert space admits to depict both possible effects and their origin.

Also in this case we determine the bound $\epsilon(w)$ with the two methods. 
We first solve problem~(\ref{supinf}), consisting in 
numerically reaching the supremum of the infimum.
Then, we apply the witness restrictions directly to the
pure states (see Eq.~\eqref{eq:pureproblem}). 
Here, this leads to
\beq
q_{{\rm min}}(p,\phi)=\bigfrac{p+2\omega^2\cos^2\phi\pm 2 |\omega\cos\phi|\sqrt{\omega^2\cos^2\phi+p-p^2}}{1+4\omega^2\cos^2\phi}\label{solution}
\eeq
where $\omega\cos\phi<0$ must be imposed, and the infimum is then obtained
for the minus sign in \eqref{solution}.

\begin{figure}
\begin{center}
  \includegraphics[width=0.44\textwidth]{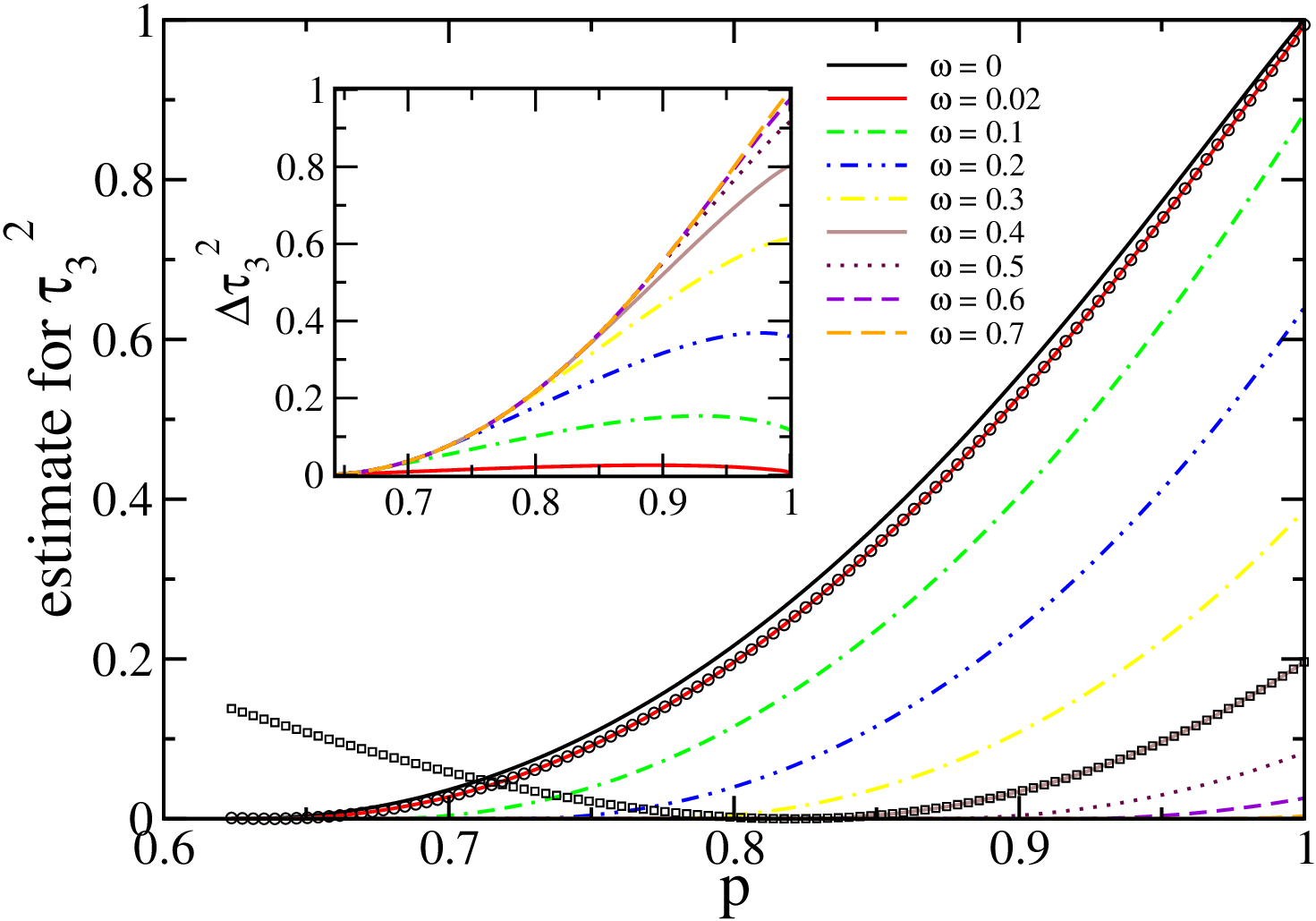}\\
\vspace*{11mm}
  \includegraphics[width=0.44\textwidth]{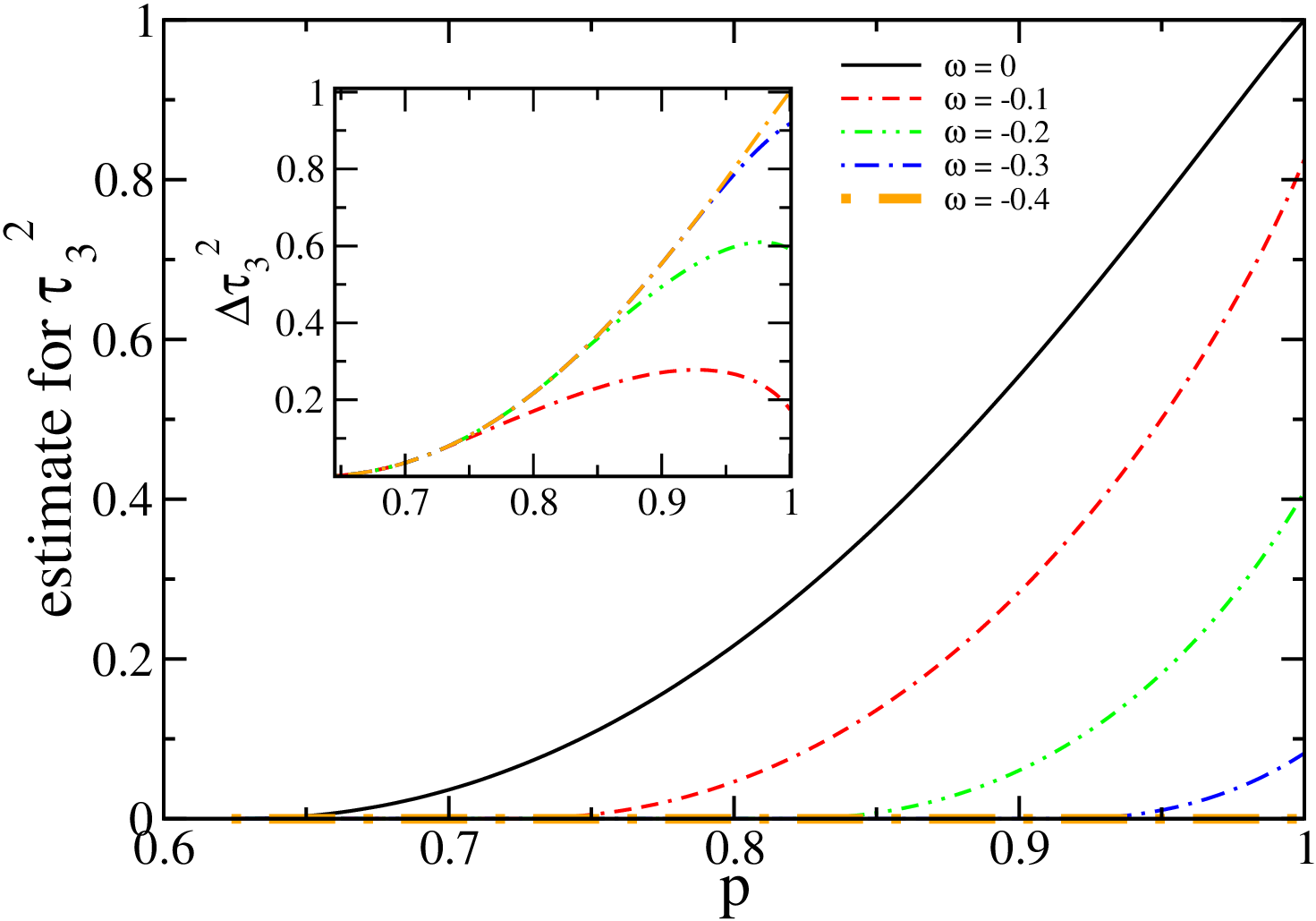}
\end{center}
\caption{(Color online) The estimate for $\tau_3^2$ from an off-diagonal witness for
various relative weights $\omega:=\alpha/\gamma$ of the off-diagonal part.
The insets show the deviation $\Delta \tau_3^2$ to the exact convex roof
for $\tau_3^2$.
Increasing $\omega$ the estimate quickly decreases and finally 
becomes the trivial bound, which is reached well before $\omega=\infty$,
which means a purely off-diagonal witness. This is only due to the special 
choice of basis we did: the off-diagonal part does not detect neither 
GHZ nor W state. The {\em lower panel} shows that negative $\omega$ is more 
destructive.
The circles and squares indicate the solution obtained from setting the 
part proportional to $r$ of the functional to zero; they lie on top
of the numerical curves.
\label{fig:tau3estimate-offdiagonal}
}
\end{figure}

In the second case, the lower bound of the three-tangle is then 
obtained as the function convex hull
of $\min_\phi\tau_3(q_{min}(p,\phi),\phi)$. For $\tau_3^2$, 
this curve is indicated with
circles in the left panel of Fig.~\ref{fig:tau3estimate-offdiagonal} 
together with the result from the numerical approach;  
their agreement is perfect~\footnotemark[\value{footnote}].

\subsection{Unrestricted optimization}
\label{sec:unrestricted}

\begin{figure}
\begin{center}
  \includegraphics[width=0.44\textwidth]{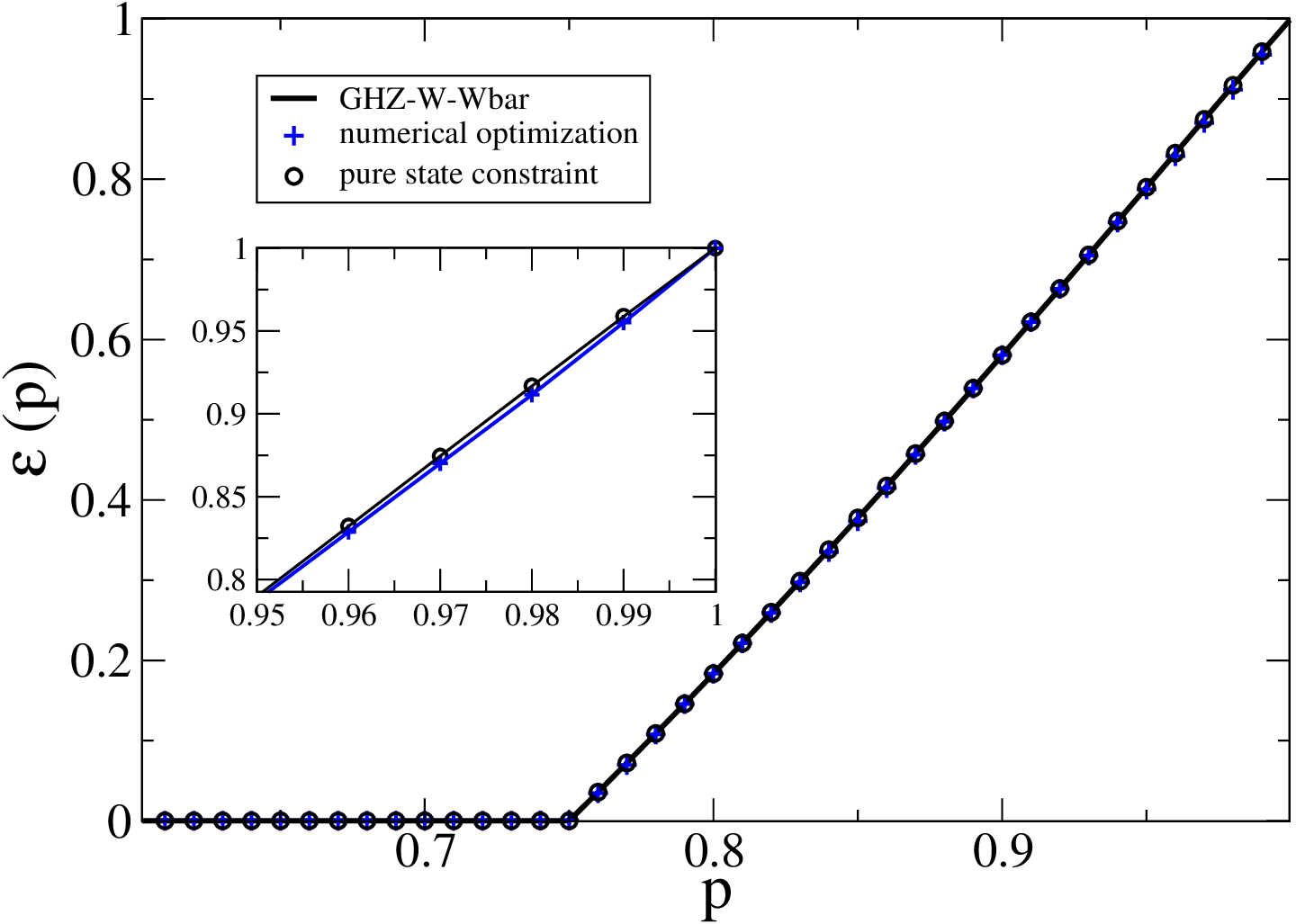}
		      \end{center}
 \caption{(Color online)
The unrestricted optimized lower bound for the
three-tangle $\tau_3$ is shown for a given GHZ fidelity $p$.
This captures also the GHZ-identity mixture. 
Two different methods are compared: full scale numerical optimization 
and the optimization 
restricted to symmetric states satisfying the constraint given by 
the GHZ witness $\frac{3}{4} \id - \pi_{\rm GHZ}$, which fixes the value of $p$.
Both methods lead to the same result (modulo taking the 
function convex hull in the latter method), as it should be.
{\em Inset}\/: close to $p=1$ the curve obtained by implying the witness 
constraint directly on the pure states is seen to lie above the curve 
resulting from numerical optimization; in this region, the pure state curve is 
not convex. The function convex hull agrees with the curve from
the full scale optimization.
}
  \label{fig:fidelity}
\end{figure}

Now let us consider the situation relevant for most experiments.
A state $\varrho$ is created and $w=\tr[\varrho {\cal W}_{\text{GHZ}}]$ is measured
\footnote{
We consider the GHZ-witness because the GHZ-state is most frequently 
aimed at in experiments \cite{Leibfried,Neumann}.
Generally, any witness based onto a projector of a state of the 
GHZ-class as defined in~\cite{Duer00} can lead to a bound $\epsilon>0$.
}.
From this value we would like to obtain an estimate of 
the tangle. This corresponds to obtaining a bound on the three-tangle
given the fidelity of the state with respect to the GHZ-state,
$p=\bra{\rm GHZ}\rho\ket{\rm GHZ}$, since $w=\alpha-p$.
We have solved the problem \eqref{supinf} for this case
numerically. The results are plotted in Fig.~\ref{fig:fidelity}.
In Table~I, the corresponding bound on the
three-tangle are listed for experiments where $p$ has been measured.
It is a curious fact that the results agree with 
those obtained by solving~(\ref{eq:pureproblem}), where the 
optimization was performed only over states which are 
symmetric under the exchange of particles. We add some more comments 
on this curious fact in Section~\ref{sec:symmetry}.

\begin{table}
\begin{center}
\begin{tabular}{|l|l|l|}
\hline
Experiment & GHZ-fidelity $p$ & $\epsilon(p)$ \\
\hline\hline
ion trap~\cite{Leibfried} & $0.86\pm 0.03$ & $0.42\pm 0.12$\\
\hline
diamond centers~\cite{Neumann} & $0.87\pm 0.06$ & $0.46\pm 0.24$ \\
\hline
ion trap~\cite{Roos} & $0.979\pm 0.002$ & $0.914 \pm 0.009$ \\
\hline
\end{tabular}
\caption{(Color online) Lower bounds on $\tau_3$ for recent experiments where the estimated $p$ 
exceeded $3/4$. The used data is the same as that used for Fig.~\ref{fig:fidelity}.}
\end{center}
\end{table}

The problem restricted on pure states 
amounts to finding the steepest descent of the 
three-tangle when the GHZ state is superposed with some other state out of 
the orthogonal complement, which is numerically found to be
reached using only the states $W$ and its bit-flipped form, $\bar{W}$.
The value zero is assumed for the weight $p=3/4$ of the GHZ state.
For GHZ-states mixed with white noise 
\beq
	\gamma \pi_{\rm GHZ}+(1-\gamma)\id/8, \quad \gamma\in\ [0,1],
	\label{eq:noisyGHZ}
\eeq
which we will refer to as {\em noisy GHZ states},
this leads to zero three-tangle at $\gamma=5/7$.
This coincides of course with the known bound for GHZ-entanglement in 
these states: by construction, if $p>3/4$ they are 
detected by the optimal GHZ witness
$W^{opt}_{\rm GHZ}=\frac{3}{4}\id-\pi_{\rm GHZ}$ (see Ref.~\cite{Acinbe}). 
Since this is a lower bound, and we do not know whether this bound is tight 
for the noisy GHZ states, we give an upper bound for its three-tangle 
at $p=3/4$.

To this end, a decomposition of the noisy GHZ state must be constructed,
and we use our result that the minimum value for the three-tangle
is assumed for symmetric states containing $W$ and $\bar{W}$ alone and 
with equal weight.
Imprinting local phases, we can induce relative phases for the $W$ and the 
$\bar{W}$ state, that are mutually inverse. 
When keeping the phase of the GHZ state fixed and
averaging over these relative phases, this leads almost
to a state of the form~(\ref{eq:noisyGHZ}),
but with the orthogonal GHZ state with relative minus sign, GHZ$_-$, missing.
Convexity of the convex-roof applied to the resulting decomposition leads 
to an upper bound of $1/9$ 
for the three-tangle of the noisy GHZ state~(\ref{eq:noisyGHZ})
with GHZ weight $p=3/4$.

An interesting question to ask here is for the full convex set of states with 
zero three-tangle and GHZ weight $3/4$. 
If the interior of this polytope contained a state of the form~(\ref{eq:noisyGHZ}),
this would mean that the lower bound was even tight for these states. 
Otherwise the bound is not tight and the threshold weight for GHZ-entanglement
(as measured by the three-tangle) would definitely be smaller
than $3/4$.

A lower bound $p_{\rm sub}=1/3$ and $\gamma_{\rm sub}=5/21$ for the weight of GHZ in a state~(\ref{eq:noisyGHZ})
without three-tangle can be obtained from a 
phase average of states of the form
$$
\gamma(p_0\pi_{\rm GHZ}+(1-p_0)\pi_{{\rm W}_\alpha})
+\frac{1-\gamma}{2}(\pi_{{\rm GHZ}_-}+\pi_{\bar{\rm W}_\alpha})
$$
which are known to have zero three-tangle~\cite{EOSU}. 
The index $\alpha$ indicates that
relative phases have been introduced into the state by means of local
phases.

\subsection{The effect of more witness constraints}
\label{sec:morewits}

If the expectation values of more than one witness operator has been measured,
then the class of states over which the optimization is performed
is restricted. 
For instance, the  problem discussed in Section~\ref{sec:restricted}
is recovered if $p=\langle \pi_{\rm GHZ} \rangle$ and $\tilde p=\langle \pi_{\rm W}\rangle$ are
known and in addition $p+\tilde p=1$ holds. In Ref.~\cite{Bourennane04},
$p$ and $\tilde p$ have been measured with a setup intended to produce a W-state.
As expected, the lower bound on $\tau_3$ obtained from the data is equal to zero.

Note that if the pure state problem~(\ref{eq:pureproblem}) returns zero
as an estimate, the final lower bound is also zero, since the function
convex hull is smaller or equal to the original function.

\section{Further observations}
\label{sec:observations}

\subsection{Restriction to pure states}\label{pure-cond}

The problem~(\ref{supinf}) can be written as
\beq
\sup_{\vec{r}}\inf_{\vec{x}}\left\{\vec{r} (\vec{w}-\vec{W}(\vec{x}))+E(\vec{x})\right\}\; ,
\eeq
where $\vec{x}$ is assumed to be a minimal set of parameters that uniquely
describes the pure states of the system.
It is worth mentioning that due to $\nabla_{\vec{r}}\nabla_{\vec{r}}\{\dots\}=0$,
the order of the extrema is relevant.
Therefore, we consider at first $\nabla_{\vec{x}}\{\dots\}=0$,
which leads to $\vec{r}\cdot\vec{W}'(\vec{x})-E'(\vec{x})$. 
If $\vec{W}'$ is invertible in a neighborhood of the 
optimal point this is formally equivalent to
\beq
	\vec{r}=\nabla_{\vec{W}}E\; ,
\eeq
that is, the vector of Lagrange multipliers equals the gradient of
the entanglement measure with respect to the witness values.
The consecutive supremum in $\vec{r}$, inserting this condition, leads
to 
\beq
	\vec{W}(\vec{x})=\vec{w}, 
\eeq
{\em i.e.}, the pure states themselves already
satisfy the experimental witness contraint.

A remark is in order: the number of witness constraints will typically
be significantly smaller than the number of parameters describing the states.
Then, for each set $\vec{w}$ of witness constraints, there will be a 
corresponding submanifold in the parameter space $\vec{x}$. 
In order that $\vec{W}'$ can be invertible,
we have to take the infimum of $E$ in that manifold and call it $E_c(\vec{w})$, 
which is the characteristic value of $E$ as introduced in~\cite{KENNLINIE},
and we can henceforth use $\vec{w}$ as the parameters describing those states
with minimal value for the entanglement measure subject to the
contraint $\bra{\psi}\vec{\cal W}\ket{\psi}=\vec{w}$.  

For positive semidefinite second derivative we have an infimum;
after straight forward algebra, this leads to the condition 
that the matrix with entries
\beq
\frac{\partial^2 E_c}{\partial w_l \partial w_m}-\frac{\partial E_c}{\partial {w_k}}  \frac{\partial^2 w_k}{\partial w_l \partial w_m} 
= \frac{\partial^2 E_c}{\partial w_l\partial w_m}
\label{eq:cond1}
\eeq
be positive definite.
This requires convexity of $E_c$. The equality holds 
where $\vec{W}'$ is invertible and the witness constraints are independent, 
i.e. we assumed $\partial_{w_l} w_k=\delta_{kl}$.
Wherever $E_c(\vec{w})$ is not convex,
the infimum will be assumed at the boundary of some interval, and the function 
convex hull will have to be taken at the end.

Interestingly, invertibility of $\vec{W}'$ and 
$\frac{\partial^2 E_c}{\partial w_l\partial w_m}$, when inserted into
the second derivative w.r.t. $\vec{r}$ leads to
$$
\left. \nabla_{\vec{r}}\nabla_{\vec{r}}(\dots)\right|_{\rm extrema\ conditions}=
-\Big(\nabla_{\vec{w}}^2 E_c(\vec{w})\Big)^{-1}\; ,
$$
which should be negative semidefinite, since we maximize with respect 
to ${\vec r}$.
This is consistent with our above result that $\nabla_{\vec{w}}^2 E_c(\vec{w})$
had to be positive definite.

This analysis generalizes the explicit calculation for a single witness
performed in Section~\ref{single-wit}. It underpins the deep connection
to the concept of characteristic curves as proposed in Ref.~\cite{KENNLINIE}.
In the presence of non-convex regions in $E_c(\vec{w})$ the extrema are
assumed at the boundary $\partial I_{\vec{w}}$ of some interval 
$I_{\vec{w}}$ in $\vec{w}$-space; in complete
analogy to the discussions on optimal decompositions in 
Refs~\cite{LOSU, EOSU,KENNLINIE} this means that no pure state satisfying
the witness conditions leads to the lower bound for this case. 
The lower bound is then
achieved by mixed states made of some pure states in $\partial I_{\vec{w}}$,
and the lower bound is affine within the whole interval $I_{\vec{w}}$.
For a single witness, this boundary consists of precisely two states and
then prescribes a mixed state solution of the problem.  

\subsection{Simplification due to symmetry}
\label{sec:symmetry}

The problem also simplifies, when both the entanglement measure to be estimated
and the observables measured in the experiment have a 
common symmetry~\footnote{
For similar arguments see Refs~\cite{TerhalPRL00,VollbrechtPRA02,BenattiJMP03,Eisert07}
and exercise 4.4 from~\cite{BoydBook04} by P. Parrillo.
}.
Let us assume that both $E(\rho)$ and $\trace{\rho {\cal W}_k}$ 
are invariant under $\rho\to \hat{Q}_j(\rho)\equiv Q_j\rho Q_j^\dagger$,
where the operators $Q_j$ form a group ${\cal G}$ with $J$ elements.
The symmetry group encountered later on
is the symmetric group $S_n$ of qubit permutations; 
therefore we assume that $J$ is finite in what follows.  
The result is more general and also applies to compact continuous groups. 
Let us define $\bar\rho=\frac{1}{J}\sum_{j=1}^J \hat Q_j(\rho)$ given a 
state $\rho$. Clearly, $\hat Q_j(\bar \rho)=\bar \rho$ holds, and then 
convexity and invariance of $E$ with respect to $\cal J$ imply 
$E(\bar \rho)\le E(\rho)$. Note that also
$\bar\rho$ fulfills the experimental constraints $w_k=\trace{{\cal W}_k\bar\rho}$ 
since also the witnesses ${\cal W}_k$ are assumed to be $\cal G$ invariant. 
Hence we can restrict ourselves to optimize over
density matrices which are invariant under the action of 
the symmetry group. 

However, this does not imply that we can 
restrict the optimization over pure states~(\ref{supinf})
to pure states with this symmetry, since the optimum of the 
convex roof for a symmetric states $\bar \rho$ 
is not necessarily taken for symmetric pure states.
To see this, assume that the problem~(\ref{eq:pureproblem})
gives the correct minimum of the problem~(\ref{supinf}) 
for a state $\ket{\psi_{\rm min}}$.
Then the symmetric mixed state
$
\bar\rho_{\rm min}:=\frac{1}{J}\sum_{j=1}^J \hat Q_j(\ket{\psi_{\rm min}})
$
also satisfies the constraints $\tr \bar{\rho}_{min} {\cal W}_k=w_k$
and fulfills $E(\bar\rho_{\rm min})=E(\ket{\psi_{\rm min}})$.
In general, for every $\ket{\psi_{\rm min}}$ that minimizes the 
problem \eqref{eq:pureproblem} we find 
$E(\bar\rho)\le E(\ket{\psi_{\rm min}})$ for (without loss of
generality) $\cal G$ symmetric states $\bar\rho$. Equality holds
{\em iff} all states in the optimal decomposition of $\bar\rho$
minimize \eqref{eq:pureproblem} separately. 
If we now assume that a $Q_{j_0}\in\cal G$ exists such that 
$Q_{j_0}\ket{\psi_{\rm min}}\neq \ket{\psi_{\rm min}}$, i.e. that $\ket{\psi_{\rm min}}$ 
is not $\cal G$ symmetric, then the ${\cal G}$-orbit of all minimal pure states
constitutes a set of optimal decomposition vectors of 
a flat roof~\footnote{It is a very peculiar
optimal decomposition; namely, where all decomposition states have the 
same value for the entanglement measure $E$. Such a singular situation 
is a general feature of sufficiently simple entanglement measures, 
as the concurrence~\cite{Wootters98}, and has interesting consequences. 
Particularly interesting among them is that taking a function of the measure 
and taking the convex roof commute in this case~\cite{KENNLINIE}.}.

The main problem we encounter in trying to extend the above proof for 
symmetric mixed states to symmetric pure states is
that the coherent symmetrization $\sum_j \sqrt{q_j} Q_j\ket{\psi_{\rm min}}$
will typically not satisfy the constraint, unless the witnesses have no
off-diagonal matrix elements for the $Q_j\ket{\psi_{\rm min}}$, i.e. unless 
$\bra{\psi_{\rm min}}Q_j^\dagger {\cal W}_k Q_l\ket{\psi_{\rm min}}\propto \delta_{jl}$.
If this condition would be satisfied, then - since all decomposition 
states of $\bar{\rho}_{\rm min}$ are superpositions of the $Q_j \ket{\psi_{\rm min}}$ 
- two cases would occur: 
\begin{itemize}
\item[a)] $E(\sum_j \sqrt{q_j} Q_j\ket{\psi_{\rm min}})\equiv E(\psi_{\rm min})$ for all $q_j$; 
$\sum_j q_j=1$.
\item[b)] A state $\ket{\tilde{\psi}}$ exists in the range of ${\cal G} \ket{\psi_{\rm min}}$ such that
          $E(\tilde{\psi})<E(\psi_{\rm min})$.
\end{itemize}
From a) we conclude that also the $\cal G$ symmetric state 
$\ket{\psi_{\rm symm}}:=\frac{1}{\sqrt{J}}\sum_{i=1}^J Q_j \ket{\psi_{\rm min}}$
is a solution of the problem~\eqref{eq:pureproblem} by virtue of the 
orthogonality condition 
$\bra{\psi_{\rm min}}Q_j^\dagger {\cal W}_k Q_l\ket{\psi_{\rm min}}\propto \delta_{jl}$.
The alternative b) is a contradiction to the initial assumption that 
$\ket{\psi_{\rm min}}$ solves~\eqref{eq:pureproblem}.
This would complete the proof that we can restrict ourselves to optimize over
$G$-invariant pure states. However, the orthogonality condition we used
in the proof seems a rather stringent at first sight;
furthermore, in order to test it one would need to find the $\ket{\psi_{\rm min}}$
first, which amounts to solving the problem without imposing the symmetry
argument. We leave this discussion for future research. Curiously,
in the cases considered in this work, the minizing pure states
turned out to be symmetric.

\section{Conclusions}
\label{sec:conclusions}

We have analyzed the method proposed in Refs.~\cite{Guehne07,Eisert07}
for obtaining a lower bound for entanglement measures from 
expectation values of witness operators. To this end, we have applied
the latter method to the convex-roof extended
three-tangle of mixed three qubit states. 
As a first general result we show that a solution to the problem is obtained,
when the constraints from the experimental
knowledge about the quantum state are imposed on pure states
instead of mixed states. The function convex hull of the
thus constructed curve gives the solution to the full problem.
This elegant simplification can be seen as a corrolary to
the results in Refs.~\cite{Guehne07,Eisert07}, 
that has not been noticed before.
Furthermore, this result highlights a close
relation to lower bounds obtained from so-called 
{\em characteristic curves}~\cite{KENNLINIE}.

We applied both the full and pure state approach to 
a simplified situation,
where the state is assumed to be of the form~(\ref{eq:teststate}), 
in order to explicitly demonstrate the 
Legendre transformation at work, using three different witnesses.
This situation corresponds to an experiment with dominating 
systematic errors: in our example, the only 
ocurring error consists in the admixture of a W state.
This analysis is useful, because it also clearly points out the 
need for a convexification after the related pure-state 
problem~(\ref{eq:pureproblem}) has been solved.
Interestingly, the thus obtained bound coincides with the exact value 
of the three-tangle as obtained in Ref.~\cite{LOSU}.
Note that this restricted approach can be generally applied if a bound on the 
entanglement of a family of states is desired.
In addition, we have analyzed off-diagonal elements 
in the witness for the same 
setting, in order to mimick the effect of not properly chosen witness 
operators. The main effect is to reduce the
estimated lower bound as compared to a perfect witness, even for 
a pure state at hand. 

We then considered the experimentally relevant case where the 
fidelity of the produced state with respect to the GHZ state 
has been measured, and no prior information about the state is assumed 
to be available.
The result is plotted in Fig.~\ref{fig:fidelity} and 
has been used to estimate the three-tangle produced in recent 
experiments, summarized in Table~I. 
We only considered experiments where 
the fidelity $\bra{{\rm GHZ}}\rho_{\rm exp}\ket{\rm GHZ}$
exceeds $3/4$, since a lower fidelity is compatible with 
a state with no three-tangle.
As to be expected, the witness bound $\gamma=5/7$ 
for having three-tangle in a GHZ-identity mixture, is reproduced.

It is a curious observation that the lower bound seems to be obtained
using permutation symmetric pure states only. 
Although the effect of common
symmetries of witness and entanglement measure - as we have here - 
is directly reflected in mixed state solutions of the optimization problem, 
the same conclusion does not seem to be straight forward for pure state 
solutions.

\acknowledgments
We acknowledge useful discussions with J. Eisert, 
O. G\"uhne, J. Siewert and H. Wunderlich.

\end{document}